\documentclass[%
 reprint,
superscriptaddress,
 amsmath,amssymb,
aps,
]{revtex4-1}

\usepackage{esint}
\usepackage{graphicx}
\usepackage{dcolumn}
\usepackage{bm}
\usepackage{lipsum}
\usepackage{color, soul}
\usepackage{enumitem}
\usepackage{ulem}

\begin{document}

\title{Backflipping motion of air bubbles colliding with a tilted wall}

\author{Alireza Hooshanginejad}
 \affiliation{%
 Department of Biological and Environmental Engineering, Cornell University, Ithaca, New York, USA.
}
 \affiliation{%
 School of Engineering, Brown University, Providence, Rhode Island, USA.
}

\author{Anuj Baskota}
 \affiliation{%
 Department of Biological and Environmental Engineering, Cornell University, Ithaca, New York, USA.
}

\author{Sunghwan Jung}%
 \email{sunnyjsh@cornell.edu}
\affiliation{%
Department of Biological and Environmental Engineering, Cornell University, Ithaca, New York, USA.
}%


\date{\today}

\begin{abstract}
Oblique collision of solid particles with surfaces has been a topic of extensive study in Newtonian mechanics, which also explains the motion of bubbles and droplets to some extent. Here, we observe that air bubbles exhibit a \textit{backflipping} behavior when they collide with a tilted surface. Our experiments reveal that bubbles with radii 0.6-0.7 mm undergo backflipping when they collide with surfaces at an angle of up to $15^{\rm{o}}$ with the strongest backflipping at $3^{\rm{o}}$. Particle image velocimetry reveals that the backflipping behavior is caused by wake-induced circulation around the bubble, which applies a lift force on the bubble. We develop a theoretical model that incorporates potential flow theory to characterize the circulation caused by the interaction between the bouncing bubble and its wake. The theoretical results are in good agreement with the experiments confirming the key role of the wake-induced lift force in backflipping. Finally, we show that the backflipping behavior of air bubbles can be leveraged for sustainable cleaning of a biological surface coated with a protein solution.  
\end{abstract}
\maketitle

The laws of collision proposed by Newton and other scientists of his era including Wallis, Wren, and Huygens provide the foundation for our understanding of {\it{elastic collision}} \cite{Birch1756}. In his famous book, {\it{the Mathematical Principles of Natural Philosophy}}, Newton discussed about the oblique collision of two spherical bodies that ``if bodies are moving in different right lines impinging obliquely one upon the other, ... the parallel motions are to be retained the same after reflexion as before \cite{Newton1729}.'' According to Newton's law of collision, during an oblique elastic collision between a rigid ball and a solid surface, the angle of incidence equals the angle of reflection \cite{Birch1756}. However, this is no longer true for inelastic collisions of an elastic ball impinging obliquely on a smooth surface \cite{Garwin1969, Kharaz2001}. Despite the differences in angles of incidence and reflection during inelastic collisions, the ball still retains its motion along the parallel direction during reflection as long as no spinning is enforced before collision \cite{Penner2002,Vollmer2011,Cross2015,Cross2018}.

Collision of an air bubble with a solid surface is considered as an inelastic collision due to the high deformation of the bubble during collision \cite{Zenit2009}. While the dynamics of air bubbles colliding with a flat surface have been studied extensively \cite{krasowska2007kinetics,Zenit2009,klaseboer2014force,manica2015force,zawala2016immortal,zawala2016influence,Wang2017}, there are only few studies on air bubbles colliding with a tilted surface \cite{maxworthy1991bubble,tsao1997observations,podvin2008model,norman2005dynamics,debisschop2002bubble,Barbosa2019}. Oblique collision of air bubbles with tilted surfaces in an aqueous medium is of great recent importance due to the urgent need for sustainable cleaning methods in the sanitization of agricultural produce to reduce foodborne illnesses \cite{Sunny1,lee2018cavitation,Esmaili2019,Hooshanginejad2022,hooshanginejad2022removing}. Although in the case of large surfaces, the shear stress associated with steady sliding of bubbles determines the average shear force \cite{hooshanginejad2022removing}, the magnitude of the shear force associated with the transient bouncing regime is much larger than the shear force during steady sliding \cite{Esmaili2019}. Hence, bouncing bubbles offer a better cleaning outcome for small contaminated areas. However, the downside of using the bouncing phase in cleaning is the spatial gaps between the collision points which leave the surface intact \cite{hooshanginejad2022removing}. Here, we report an intriguing behavior by an air bubble of a certain size range, where it bounces backward after its first collision with a tilted wall. We show that this novel backward motion of the bubble does not follow Newton's laws of collision, which is due to the hydrodynamic interaction between the bouncing bubble and its rising wake. While few studies have shown the transverse motion of a bubble rising next to a vertical wall due to a transverse lift force \cite{TAKEMURA2002,takemura2003transverse}, the backward motion of bubbles along a tilted surface has never been explored to the best of our knowledge.

\begin{figure*}
\centerline{
\includegraphics[scale=0.9]{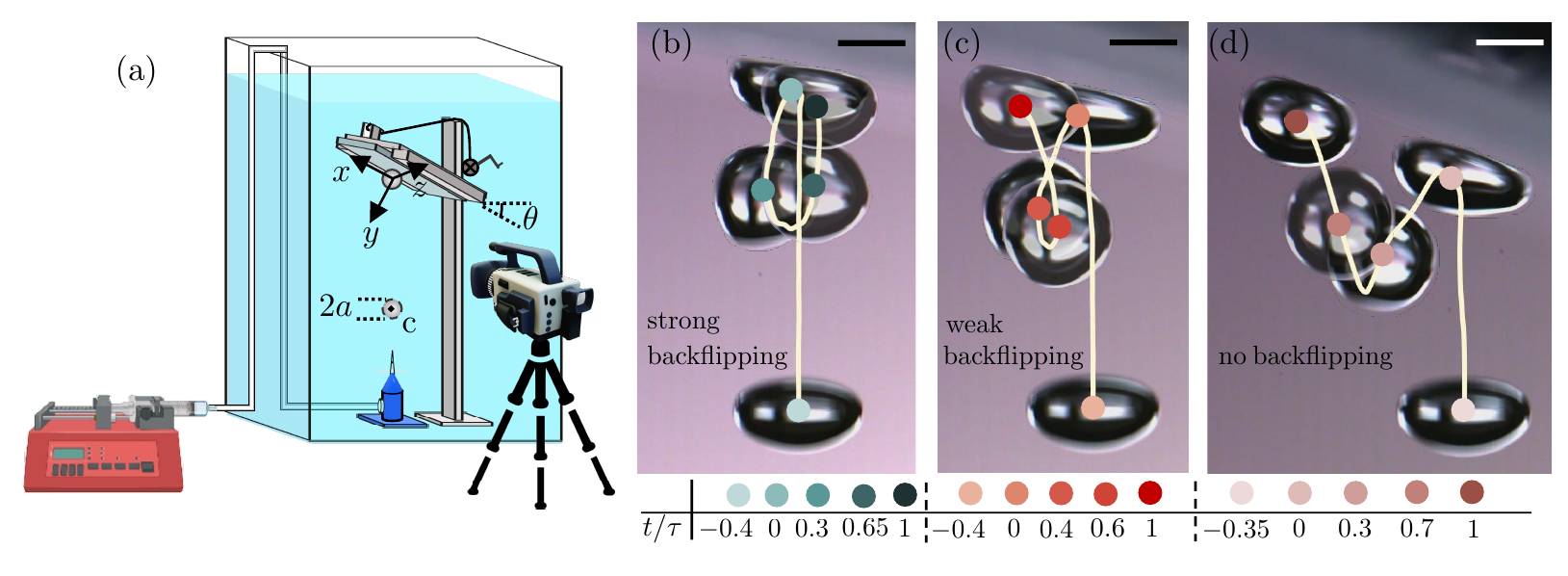}
}
\caption{\label{fig:1} (a) The schematic of the experimental apparatus. (b)  At $\theta=5^{\rm{o}}$, a bubble with radius $a=0.7$ mm bounces backward and collides the surface below the first collision point. (c) At $\theta=10^{\rm{o}}$, the bubble bounces backward, but ultimately collides at a point above the first collision point. (d) At $\theta=20^{\rm{o}}$, the bubble no longer undergoes backflipping. The corresponding videos are included in Ref. \cite{SupplementalMaterial}. 
}
\end{figure*}

We perform experiments in a tank filled with deionized water to visualize the bubble motion while colliding with a tilted smooth surface. We place a custom-made needle at the bottom of the tank connecting the needle to a syringe pump (New Era) through a 3D-printed mount and tubing as depicted in Fig. \ref{fig:1} (a). The needle is made to generate bubbles of uniform radius at low pump flow rates with a standard deviation of less than 0.02 mm. We then use a 3D-printed tower that has a manual pulley design to hold a clean glass slide at the desired inclined angle, $\theta$, on top of the needle as shown in Fig. \ref{fig:1} (a). The distance between the needle and the lowest edge of the glass slide is approximately 12 cm to ensure that the bubbles reach their terminal velocity before collision. We use a high-speed camera (Photron SA-Z), a macro lens (Nikkor 105 mm), an extension bellows (Nikon PB-6), and a 10x magnifying lens (Neewer) to record the bubble behavior near the collision zone. We test three different bubble radii, $a \simeq 0.6$, $0.65$, and $0.7$ mm. We note that bubbles of the current size range have an oblate spheroid shape during the rising phase, but retain a spherical shape after the collision. We also note that all bubbles rise along a straight path as the associated Reynolds number (i.e. ${\rm{Re}}=\rho U_{\rm{T}} a/\mu\sim180-250$) is below the threshold of path instability (i.e. $\rm{Re_{cr}}\sim 300-400$) \cite{clift2005bubbles}. Here, $\rho$ denotes the density of water, $\mu$ the dynamic viscosity of water, and $U_{\rm{T}}$ the terminal velocity of the bubble. In addition to different bubble sizes, we test different $\theta$, ranging from $0^{\rm{o}}$ to $20^{\rm{o}}$.

Figure \ref{fig:1}(a) shows different trajectories of an air bubble with $a=0.7$ mm from the first collision at $t= t_0$ to the second collision at $t=t_{\rm{1}}$. We define a dimensionless time, $\tau=(t-t_0)/(t_{\rm{1}}-t_0)$, normalized by its period between the two bounces (i.e., $t_{\rm{1}}-t_0$). 
At $\theta=0^{\rm{o}}$, the bubble bounces normal to the surface (video 1 in Ref. \cite{SupplementalMaterial}). As we tilt the surface at a low angle $\theta$, the bubble initially bounces forward like an elastic collision, but then reverses and moves backward until it collides with the surface at a location behind the first collision spot, as shown in Fig. \ref{fig:1}(b) for $\theta=5^{\rm{o}}$ (video 2 in Ref. \cite{SupplementalMaterial}). We herein define the backward motion of the bubble as \textit{backflipping} as shown in Fig. \ref{fig:1}(b). As we increase $\theta$ to $10^{\rm{o}}$, the bubble still exhibits a backflipping behavior; however, the bubble collides on a point past the first collision point, as shown in Fig. \ref{fig:1}(c) (video 3 in Ref. \cite{SupplementalMaterial}). We refer to such a backflipping motion of the bubble as \textit{weak backflipping}. Finally, at $\theta=20^{\rm{o}}$, the bubble no longer backflips as shown in Fig. \ref{fig:1}(d) (video 4 in Ref. \cite{SupplementalMaterial}). 


\begin{figure*}
\centerline{
\includegraphics[scale=0.9]{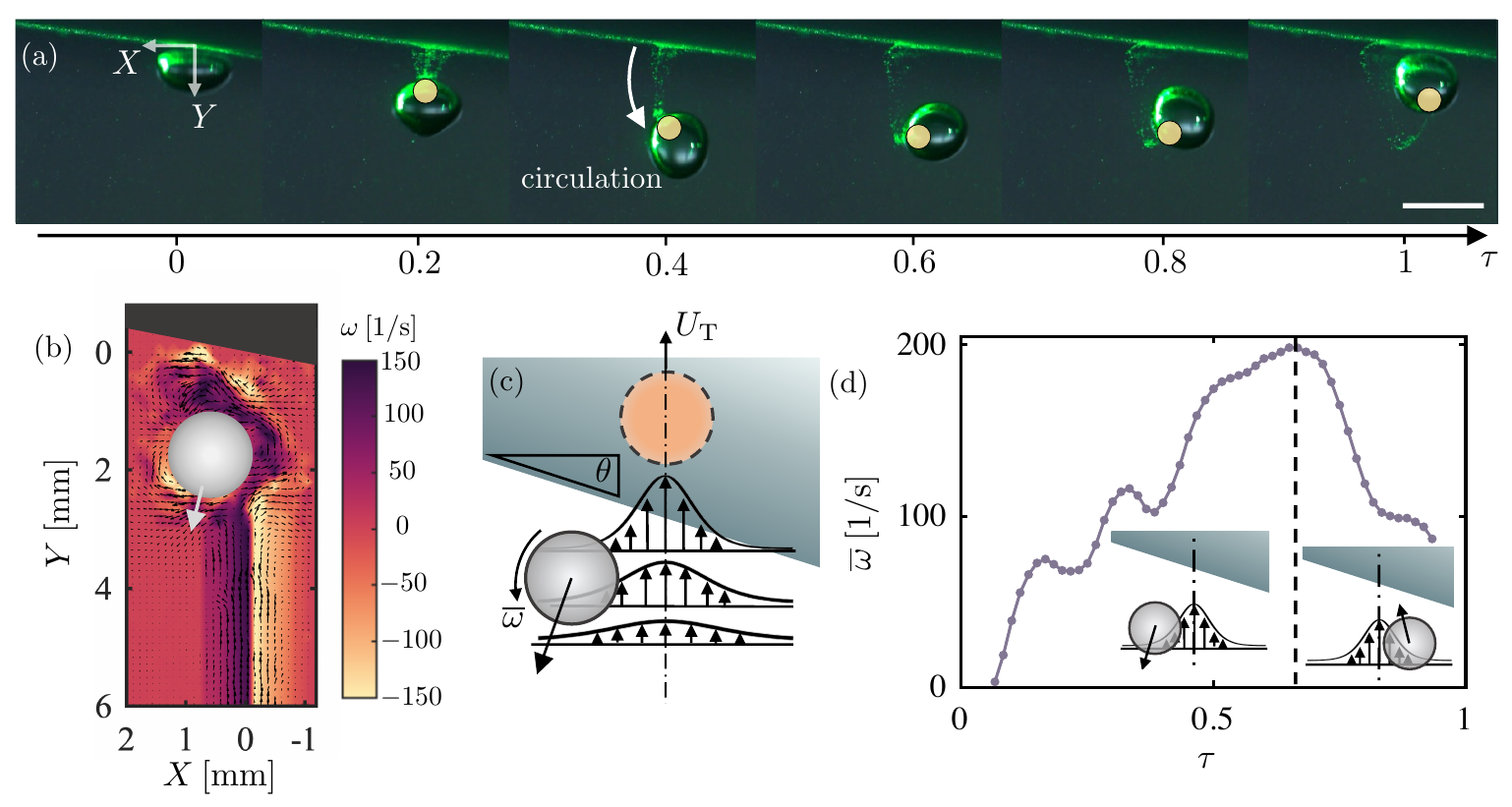}
}
\caption{\label{fig:2} (a) Time sequential images of a bouncing bubble with $a=0.7$ mm at $\theta=5^{\rm{o}}$ for $\tau=0-1$ (video 5 in Ref. \cite{SupplementalMaterial}). The yellow circle following the particles on the bubble surface indicates a counterclockwise circulation around the bubble. (b) The vorticity map from PIV around the bubble at $\tau=0.3$ for a bubble with $a=0.7$ mm and $\theta=1^{\rm{o}}$ (video 6 in \cite{SupplementalMaterial}). (c) Schematic showing the use of a reference bubble rising continuously to model the interaction between the bouncing bubble and its own rising wake. (d) The average vorticity, $\overline{\omega}$, around the bubble surface over time for $a=0.7$ mm, and $\theta=1^{\rm{o}}$.}
\end{figure*}

To understand the key flow features associated with backflipping, we visualize the flow around the collision zone by adding 10 $\mu \rm{m}$ hollow glass particles to the tank and using a 100 milliwatt laser sheet (Laserland Co.) to illuminate the bubble motion in a 2D plane. We leave the seeding particles to rest in water; therefore, some particles aggregate on the surface of the tilted glass wall. Figure \ref{fig:2}(a) shows a bubble with $a=0.7$ mm colliding with a surface at $\theta=5^{\rm{o}}$ from $\tau=0$ to $\tau=1$ (video 5 in Ref. \cite{SupplementalMaterial}). When the bubble collides with the surface, the particles on the surface detach and follow the bouncing bubble. Specifically, these particles show a signature of counterclockwise circulation around the bubble, as shown in Fig. \ref{fig:2}(a). To characterize the circulation around the bubble, we perform more rigorous particle image velocimetry (PIV) experiments. Figure \ref{fig:2}(b) shows the vorticity distribution, $\omega(X,Y)$, in a laboratory frame, $XY$, around a bouncing bubble with $a=0.7$ mm, and $\theta=1^{\rm{o}}$ at $\tau=0.25$ (video 6 in Supplementary Materials \cite{SupplementalMaterial}). The vorticity colormap of Fig. \ref{fig:2}(b) shows that the bouncing bubble moves through a shear flow structure created in the past, as illustrated in Fig. \ref{fig:2}(c). 

As the bubble approaches the surface for the first collision, it generates an axisymmetric flow in its wake. Note that we neglect any unsteadiness in the wake of the rising bubble as the current range of bubble size remains below the threshold of vortex-induced path instability \cite{Haberman1956,Duineveld1995}. The circulation caused by the shear flow in the background induces a lift force on the bubble that is always pointed towards the bubble's trajectory prior to the first collision. Figure \ref{fig:2}(d) shows that the time-averaged vorticity around the bubble surface, $\overline{\omega}(\tau)$, first increases in time, and then decays. It is worth noting that $\overline{\omega}$ starts to decay at the moment when the centroid of the bubble crosses the central axis of the rising bubble. This is due to the opposite direction of the shear flow over the bubble as depicted in the inset schematic of Fig. \ref{fig:2}(d).

\begin{figure*}
\centerline{
\includegraphics[scale=0.9]{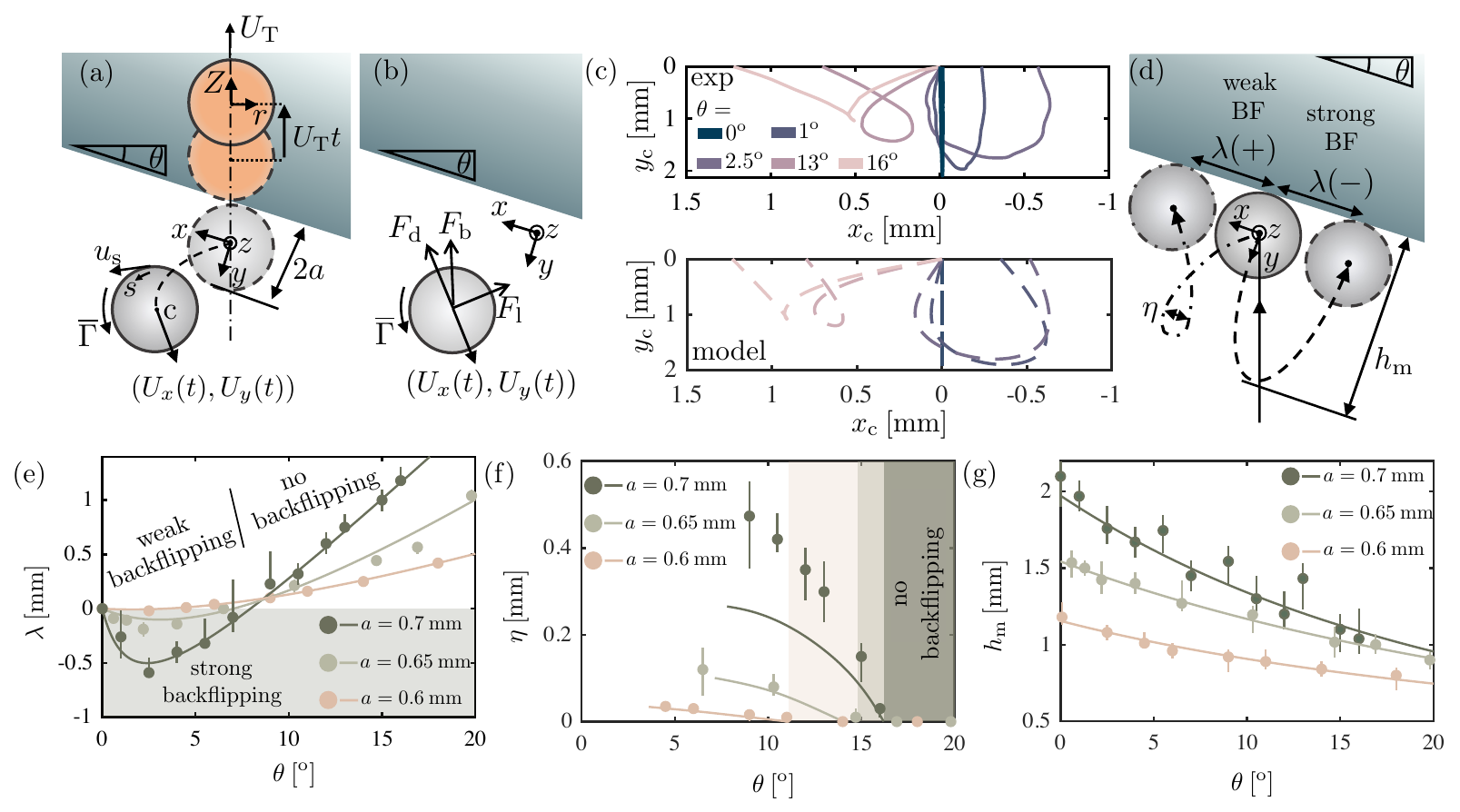}
}
\caption{\label{fig:3} (a) The schematic for the potential flow model and the freely rising image bubble. (b) The force balance schematic for the bouncing bubble. (c) The bubble trajectory from the experiments and the model for $a=0.7$ mm, and varying $\theta$. (d) The schematic showing geometrical parameters characterizing the backflipping behavior. Comparison between the experiments and the model for (e) $\lambda$ vs $\theta$, (f) $\eta$ vs $\theta$, and (g) $h_{\rm{m}}$ vs $\theta$ at varying $a$. The error bars correspond to 5 trials for each case. }
\end{figure*}

We mathematically model the wake from the free rising phase of the bubble as a potential flow. To find the wake-flow profile around the bubble at each time, we consider an image bubble after collision. The image bubble continuously rises with $U_{\rm{T}}$ behind the surface wall, hence separating from the real bubble bounced upon collision. Then, the potential flow induced by the rising image bubble represents the primary flow around the real bubble induced by its own wake during the bouncing phase, as depicted in Fig. \ref{fig:3}(a). We note that we neglect the secondary flow from the reflection of the rising wake flow. By approximating the bubble as a sphere, the velocity potential associated with the image bubble yields $\phi=-U_{\rm{T}}(a^3/2)Z/(r^2+Z^2)^{3/2}$ where $r$ and $Z$ are the radial and axial variables of a cylindrical coordinate moving with the centroid of the image bubble also shown in Fig. \ref{fig:3}(a). Therefore, the tangential velocity $u_{\rm{s}}$ along the bubble surface is given at any position of the bubble where $s$ denotes the circumferential length in the $xy$ plane. The average circulation around the bubble is then given by $\overline{\Gamma}=(1/2a)\varoiint u_{\rm{s}}dsdz$, where $z$ denotes the out-of-plane coordinate as shown in Fig. \ref{fig:3}(a). The lift force is correlated with the circulation around the bubble as $\bm{F_{\rm{l}}}=(4\pi a^3/3)C_{\rm{l}} \bm{\overline{\omega}} \times \bm{U} $ \cite{legendre1998lift} where $\bm{U}(\tau)$ denotes the velocity at the center of the bubble, $\bm{\overline{\omega}}(\tau)=\overline{\Gamma}(\tau)/\pi a^2$ denotes the average vorticity around the bubble, and $C_{\rm{l}}$ denotes the lift coefficient that is kept constant following previous studies \cite{tomiyama2002transverse,adoua2009reversal,Esmaili2019}. Further details of the potential flow model are provided in Supplementary Material \cite{SupplementalMaterial}.

We develop a simplified model for bubble motion from the first collision to the second collision that incorporates the effect of wake-induced lift force on bubble motion. Figure \ref{fig:3}(b) shows the free-body diagram for the bouncing bubble. We consider the buoyancy force, $\bm{F_{\rm{b}}}$, the drag force, $\bm{F_{\rm{d}}}$, and the lift force, $\bm{F_{\rm{l}}}$, to be the primary forces that affect the bouncing bubble. We also consider an added mass term due to accelerating or decelerating motions of the bubble. After some algebraic manipulations, the force balance in the $xy$ plane yields 
\begin{equation}
	\frac{{\rm{d}}\bm{U}}{{\rm{d}}\tau}=\bm{S}\bm{U}+\bm{B},
   	\label{Eqn: force balance ode}
\end{equation}
where $\bm{S}$ is a rank-two tensor corresponding to lift and drag forces as
\begin{equation}
\bm{S}= \frac{1} {C_{\rm{m}}}
\begin{bmatrix}
    -\frac{\pi}{4} {C_{\rm{d}}}  \mu a \rm{Re}  &  {C_{\rm{l}}} \overline{\Gamma}(\tau)      \\
    - {C_{\rm{l}}} \overline{\Gamma}(\tau)  &   -\frac{\pi}{4} {C_{\rm{d}}}  \mu a \rm{Re}     
\end{bmatrix},
\label{Eqn: S matrix}
\end{equation}
and $\bm{B}$ denotes the body force vector associated with buoyancy effects,
\begin{equation}
\bm{B}= \frac{g}{C_{\rm{m}}} 
\begin{bmatrix}
    \sin{\theta}        \\
    \cos{\theta}         
\end{bmatrix}.
   	\label{Eqn: B matrix}
\end{equation}

Here, $C_{\rm{d}}$ and $C_{\rm{m}}$ denote the drag coefficient and the added mass coefficient, respectively. The values of all coefficients are chosen following previous studies \cite{Esmaili2019,hooshanginejad2022removing}. Equation \ref{Eqn: force balance ode} is numerically solved with different values of $a$ and $\theta$. We use the initial velocity as  $\bm{U}(t=0)=(U_0\sin{2\theta},U_0\cos{2\theta})$, where $U_0$ denotes the initial speed of bouncing at $\theta=0^{\rm{o}}$ from the experiments for each bubble size. Figure \ref{fig:3}(c) shows the comparison of the bubble trajectory between the model results and the experiments for $a=0.7$ mm with different $\theta$. As indicated in Fig. \ref{fig:3}(c), the simplified model qualitatively matches the experiments in capturing strong backflipping, weak backflipping, and no backflipping behaviors. To characterize the backflipping behavior more quantitatively, we define a collision distance, $\lambda$, that denotes the distance between the second collision and the first collision locations along $x$. As demonstrated in Fig. \ref{fig:3}(d), the negative value of $\lambda$ corresponds to the strong backflipping behavior, while the positive value of $\lambda$ includes both weak backflipping and no backflipping behaviors. Figure \ref{fig:3}(e) shows $\lambda$ from theory for three different bubble sizes, $a=0.6$, 0.65, and 0.7 mm at different $\theta$ ranging from $0^{\rm{o}}$ to $20^{\rm{o}}$ in good agreement with the experiments. As shown in Fig. \ref{fig:3}(e), the backflipping behavior becomes pronounced as $\theta$ increases in the beginning and then starts to decay until the bubble transitions to the weak backflipping regime. We note that bubbles smaller than the presented range do not show any backflipping behavior as a result of their weaker wake. In addition, bubbles larger than the current range exhibit 2D or 3D rising motion before collision, which makes controlling the collision angle extremely complicated. 

\begin{figure}[b]
\centerline{
\includegraphics[scale=0.9]{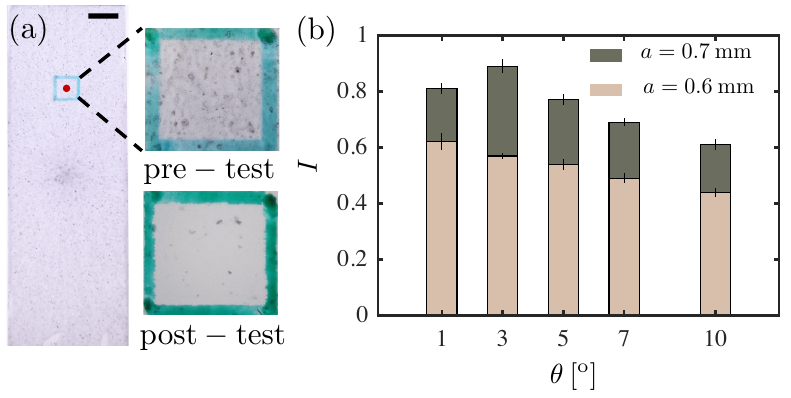}
}
\caption{\label{fig:4} (a) A sample glass slide double-coated with a protein soil solution before and after the test [$a=0.7$ mm, and $\theta=1^{\rm{o}}$]. The processing area is marked with a green box centered by the first collision point marked with the red circle. The scale bar represents 4 mm. (b) The transmission factor, $I$, for varying $\theta$. $I=1$ refers to a completely clean surface while $I=0$ refers to the original coated surface. The error bars correspond to 3 trials for each case.}
\end{figure}

The weak backflipping regime cannot be differentiated from the no backflipping regime with the parameter, $\lambda$. Hence, we define a new parameter, $\eta$, as the maximum distance traveled backward along $x$ when $\lambda>0$ as illustrated in Fig. \ref{fig:3}(d). We note that $\eta= 0$ and $\eta>0$ correspond to no backflipping and weak backflipping, respectively. Figure \ref{fig:3}(f) shows the model predictions for $\eta$ vs. $\theta$ for different $a$. Both the model and the experiments show a decaying trend for $\eta$ in the weak backflipping regime. In addition, as the bubble size decreases, both $\eta$, and the range of $\theta$ for the weak-backflipping behavior decrease. While the model predicts the threshold of the no backflipping behavior in good agreement with the experiments, we note that the $\eta$ predictions from the model are slightly smaller than the experiments for $a=0.7$ mm. This is presumably due to the secondary effect of bubble deformations, which becomes pronounced as the bubble size increases. Finally, Fig. \ref{fig:3}(g) shows the maximum bouncing height, $h_{\rm{m}}$, from the surface for both the experiments and the model in good quantitative agreement. Overall, as $\theta$ increases, $h_{\rm{m}}$ decreases as the normal component of the collision velocity scales as $U_{\rm{T}}\cos{\theta}$. 

To test the efficacy of backflipping behavior in cleaning biological surfaces, we run a series of experiments on surfaces coated with a protein solution. We first double-coat glass slides with a protein soil solution using a spin-coater \cite{Hooshanginejad2022}. We then designate an area of 4 mm$\times$4 mm, and target its center for the first collision of the bubbles. We continuously inject bubbles from the needle with a frequency of  $\sim$20 bubbles per minute for 15 minutes (i.e. $\sim$ 300 total bubbles). We use a white LED background to image the clean surface before coating, the coated surface before the test, and the surface after the test. The average values of the grayscale images corresponding to each are denoted with $I_0$, $I_1$, and $I_2$, respectively. Then, we define the normalized intensity as $I=(I_2-I_1)/(I_2-I_0)$ as the efficacy of the cleaning experiments. Figure \ref{fig:4}(b) shows that the best cleaning efficacy (i.e. maximum $I$) happens for $a=0.7$ mm, and $\theta=3^{\rm{o}}$ which coincides with the maxima in $\lambda$ shown in Fig. \ref{fig:3}(d). Further details about the cleaning test method are included in the Supplementary Material \cite{SupplementalMaterial}. 


As bubbles bounce backward in the strong backflipping regime, more surface area is exposed to bubbles and gets cleaned due to the shear forces. Therefore, the cleaning efficacy are more pronounced in the target area when the bubbles backflip. The backflipping phenomenon can be used in the design of sustainable devices for sanitizing surfaces where an array of nozzles can be used to target different points on the surface. As discussed above, the key to getting backflipping bubbles is to control the bubble size and collision angle. It is noteworthy that the wake-induced circulation discussed here is not exclusive to the current system and can have potential applications in other systems where redistributing particle aggregates on a surface is of interest.

The authors thank B. Cooke, T. J. Sheppard, and P. Xu for their help in making the spin coater and surface coatings. We also thank D. Koch, R. Zenit, and J. Magnaudet for fruitful discussions. The authors acknowledge funding from the National Science Foundation under Grants No. CBET-1919753, and No. CBET-2042740.

\bibliography{Bubble2}

\end{document}